\documentclass[prE,twocolumn]{revtex4}
\usepackage[cp1250]{inputenc}
\usepackage{color}
\usepackage{graphicx}
\usepackage{hyperref}

\begin{document}

\title{Aliens on Earth. Are reports of close encounters correct?}

\author{Pawel Sobkowicz}
\affiliation{ \url{pawelsobko@gmail.com}}

\date{April 1st, 2012}

\begin{abstract}
Popular culture (movies, SF literature) and witness accounts of close encounters with extraterrestrials provide a rather bizarre image of Aliens behavior on Earth. It is far from stereotypes of human space exploration. The reported Aliens are not missions of diplomats, scientists nor even invasion fleets; typical encounters are with lone ETs (or small groups), and involve curious behavior: abductions and experiments (often of sexual nature), cattle mutilations, localized killing and mixing in human society using various methods. Standard scientific explanations of these social memes point to influence of cultural artifacts (movies, literature) on social imagination, projection of our fears and observations of human society, and, in severe cases, psychic disorder of the involved individuals.  In this work we propose an alternate explanation, claiming that the memes might be the result of observations of actual behavior of true Aliens, who, visiting Earth behave in a way that is then reproduced by such memes. The proposal would solve, in natural way, the Fermi paradox.
\end{abstract}
\maketitle

%\pacs{}
%\keywords{Extraterrestrial civilizations; Fermi paradox; Alien abductions}


\section{Existence of extraterrestrial civilizations }
\subsection{Drake equation}
The two major problems faced today by scientists interested in extraterrestrial civilizations are: how many such civilizations exist today or existed before in the Milky Way? Provided the positive answer to the first issue, the second question is: was there, at any time and in any form, a contact between such alien civilization(s) and humanity?

A concrete answer to the first problem is clearly impossible. We know that at least one civilization (ours)  has arisen during several billions of years of the galactic history, actually quite late in the cosmic time, and after very slow start of life on Earth. But efforts to estimate the number of \textit{other} civilizations have been based mostly on educated guesses. An attempt to put some rigorousness  into these speculations is the Drake equation, expressing the frequency of civilizations through a series of more or less understood variables \cite{drake62-1,shklovskii66-1}. 
We follow here a simplified form of Drake equation, discussed by Franck et al. \citep{franck01-1}, because that work extends much beyond the limitations of the standard approach.
\begin{equation}
N_{CIV} = N^*_{MW} \cdot f_P \cdot n_{CHZ} \cdot f_L \cdot f_{CIV},
\end{equation}
where $N^*_{MW}$ is the total number of stars in the Milky Way; $f_P$ is the fraction of stars with Earth-like planets; $n_{CHZ}$ is the average number of planets suitable for the development of
life per planetary system; $f_L$ is the fraction of habitable planets where life
emerges and a full biosphere develops; $f_{CIV}$ denotes the fraction of these developing
technical civilizations. Some of these parameters are reasonably known (for example $N^*_{MW}\approx 10^{11}$) while some ($f_L$, $f_{CIV}$) are purely speculative. Franck et al. have  estimated the number of extra-solar terrestrial planets with a globally acting
biosphere as half a million. We note, however, that they have used $f_P=0.01$, i.e. a planetary system per a hundred stars. Recent report, based on increased ratio of planet detection based on microlensing \citep{cassan12-1}, puts this ratio at close to 1 (estimating number of planets as close to the number of stars in the Galaxy). Applying this correction to Franck et al. estimates gives us much larger number of Earth-like planets -- about 50 million.

Discussing temporal aspects of the Drake equation, \'{C}irkovi\'{c} \citep{cirkovic03-1}, 
stresses the importance of the longevity of civilizations.
This is especially interesting from our point of view. In this paper we are focused on possible contacts with advanced civilizations existing for a long periods of time (compared to human viewpoint), for example those existing for a million of our years. This is still short on the Galactic  
timescale (measured in billions of years) and on the scale of  evolution of main sequence stars. 
Yet, a million years is long enough to allow travel over interstellar distances. 
We note here, that we assume a strict adherence to relativity physical limitations, so that
advanced civilizations may have technical means to build practical (but slow) spaceships, but would not posses faster-than-light technologies. Therefore all hypothetical contacts would suffer from inevitable delays due to huge distances.

As we have mentioned, our complete lack of knowledge about $f_{CIV}$ makes any speculations truly vague. For the purposes of our analysis we shall \textbf{assume} that the value is high enough to allow existence of at least one (possibly a few) of advanced civilizations, within a contact time/distance horizon, during the human recorded history (past 10~000 years). 

\subsection{Fermi paradox}
`If they exist, they should be here' -- this is the gist of a paradox, reportedly stated by Enrico Fermi in 1950, and therefore bearing his name. Given enough time and resources, advanced civilizations should have reached our planet (among many others) and thus we should be able to detect their presence (current or historical). Yet, mainstream science does not confirm discovery of traces of such visitors. In fact the most popular solution to the Fermi paradox is that there are no advanced, space-faring civilizations (e.g. \citep{hart75-1,tipler80-1,brin83-1,freitas83-1,wesson90-1}).
Many of these explanations are based on an assumption that an advanced civilization, capable of interstellar travel would colonize the Galaxy directly (i.e. sending out living Aliens) or indirectly (via Von Neuman automated, replicating probes). Assuming exponential growth, despite vast scale of the Milky Way, eventually all stars and planets would be visited and/or colonized by the first sufficiently advanced civilization -- the analogy of human expansion on Earth is obvious. 

In addition to this trivial solution to the Fermi paradox, there are other solutions, assuming different characteristics of the extraterrestrial civilizations. For example they may fall into `silent' or `noncolonizing' categories \citep{hetesi06-1}. Or they may be present here, but very effectively hide this presence. For example they may be present in the Solar system and observe us, undetected, from a distance. In what follows, we shall \textbf{assume} that our planet has indeed been `discovered' by extraterrestrials, and look into ways that such assumption may be reconciled with the Fermi paradox.

\section{Aliens on Earth: popular memes and explanations\label{standardexplanations}}

\subsection{Large scale presence: global friendly contact, invasions...}
For more than a hundred years, popular culture has entertained many variants of planetary scale contacts with extraterrestrial civilizations. Most frequent are the artistic accounts of alien invasions, with Wells' \textit{War of the Worlds} as an archetype. Accounts of peaceful Contact are less numerous, most likely due to smaller narrative power. The alien invasion theme resonates very deeply with human fears: the horror of wars is well known, and the space invaders are the ideal example of stigmatized `otherness', allowing the stories to stress human unity in combat (or defeat). Variants of the invasion stories range from deadly serious and tragic, through heroic to comic, and in many (most?) cases can be `translated' into humanity `internal' narratives. Artistic visions of 
such invasion are too numerous to provide, even sketchily. On more serious nature, one can point out to \citep{mumfrey09-1} -- a `manual' preparing those who are willing to listen to the coming apocalyptic alien invasion.  

The peaceful Contact stories are often based on our imagination of what human society would do in case of interstellar exploration. Here the dominant motif if scientific/cultural expeditions, where the Contact serves increase of knowledge. Less frequent alternative is the commercial exploration (usually just transforming earthbound corporate stories to other planets). 

What is interesting from our point of view, is that although in literature and movie industry planetary scale Contact is very popular, this popularity is not shared in the `grassroot' memes of reported contacts with Aliens on Earth. While the Orson Welles radio drama has  resulted in massive panic, there are very few reports on massive presence of Aliens on our planet. One of the reasons -- especially in our `instant news' age -- is that it is hard to believe that such presence (in form of an invasion, or full scale diplomatic, scientific or commercial contact) would be unobserved. There are, of course, explanations based on conspiracy theories, but they would require not only effective cooperation between human agencies and governments but also the willingness of the extraterrestrial representatives to remain hidden.  

\subsection{Small scale phenomena: sightings, abductions, infiltration\ldots}
The situation is different for small scale Contact, where small groups of humans interact with lone Aliens or small groups of such visitors. Here the reports of supposed Third Kind encounters parallel strongly the literary and cinematographic culture. There are various recurrent motifs of such close contacts. By far the most frequent are  sightings of UFOs \citep{friedman11-1}. 
They are present in popular culture (attributed to extraterrestrial civilizations) for more than 70 years. While some cases have been explained as natural phenomena or human built artefacts, the anecdotal nature of accounts makes them hard to analyze scientifically. Advances in image manipulation technologies makes even the `hard records' doubtful. It is therefore not surprising that while in popular culture   UFOs are very much alive and present, the mainstream science gives very little attention to such reports.
We draw the attention to one of recurring themes in accounts of UFO sightings: in many cases the behavior of the UFOs is reported as erratic, at least from the human point of view.  

An example of more direct interactions with the extraterrestrials are so called `Alien abductions' -- accounts of situations where single humans or small groups were taken to extraterrestrial spaceships and (often) experimented upon. Here, the scientific research has been more detailed -- but taking entirely `earth-bound' explanation path. Most of such studies focus on psychological or sociological explanations. A `mild' study by Bullard \citep{bullard89-1} suggests that the abduction accounts are a new guise of `supernatural kidnap meme'. Other authors focus on social effects \citep{patry01-1} or psychiatric aspects \citep{bartholomew91-1,spanos93-1,banaji96-1,lynn96-1,mcnally05-1}. 
The latter frequently point out to mental disorders of the people reporting close encounters with Aliens. Generally, the approach is to look for disorders in the state of the `messengers' bringing the news of the encounter, and not on the nature and consequences of such and event.
Similar analyses related to reports of other types of activities of ETs, such as cattle mutilation, also focus on psychiatric explanations \citep{bartholomew92-1}. We find it quite interesting that in his review of various explanations of abduction reports, Eberlein \citep{eberlein01-1} actually considers the `naive-realistic' hypothesis (stating that at least some of the reports are based on true events) as one of the possible research hypotheses,   
but stresses its delusional origin due to the link with Science Fiction literature and films.

Similar influence of SF imagery may be found in the reports of Alien infiltration of human societies, whether in the form of shape-changing beings mimicking our appearance or body/mind snatchers using zombified people \citep{webb08-1,gelman10-1}. While, on the surface, the reports on observations of Aliens hiding in midst of our society seem simply to reproduce the stereotypes of novels or movies, one should bear in mind that the artistic imagery, in turn, may originate from unusual behavior of individual people. Thus classifying cause/effect may be rather difficult.

Generally, the mainstream science treats reports of current or past presence of the Aliens 
on Earth as pseudoscience. It is interesting to note that the popular audience accepts the 
accounts -- as may be judged by the sales of books related to such phenomena. The `debunking' counterparts, with a few  exceptions, do not sell so well and do not enter the `popular mindset' 
with equal appeal. One of the reasons is, of course, that the rigors of scientific research make it much more difficult and less appealing than emotional narratives and `easy' explanations \citep{sagan75-1}. This is clearly a psychological/social phenomenon with no relation to 
Alien encounters. But the virtual lack of serious investigations considering, as a working hypothesis, that some of the accounts may be based on real events is also quite interesting.    

\subsection{Conspiracy theories}
Probably the most popular explanation about the lack of verified, official confirmation of contacts with the Extraterrestrials (large or small scale) is based on conspiracy theories. Various variants of the Men in Black government cover-up exists since the very beginning of the popularity of memes about `flying saucers' in 1947. The expositions of government cover-ups are almost as popular as accounts of contacts with the Aliens \citep{frazier97-1,graham11-1}. Interestingly, such government conspiracy stories are actually stronger in societies with long tradition and practice of free speech (US, UK\ldots) than in countries where censorship of news was an actual practice in everyday life (like communist block). Perhaps an explanation is that citizens of communist state understood much better the futility of censorship effort in blocking the information. US citizens, on the other hand, could imagine a truly efficient `blackout', due to US propaganda about the all-encompassing communist regimes\ldots

While external censorship is, indeed, very hard to maintain, there remains an option of self-sustaining internal censorship, due to cultural structure of the research community. 
The avoidance of being ridiculed, the need to publish and to get research grants are important factors shaping the behavior of most scientists. Showing serious consideration of hypotheses outside the accepted worldviews may be dangerous to the career path.  
The modern way of financing and governing research channels the efforts into well established mainstream, with decreasing probability of success of controversial approaches.

\subsection{Zoo hypothesis}
A variant of the `conspiracy' explanation of the Fermi paradox is the `Zoo hypothesis', \citep{ball73-1}, according to which our planet has been set up as a galactic `preserve', with Aliens themselves declaring it a `no go' zone. This assumes an efficient control on the part of the Aliens (against their own trespassers or accidental disclosures). 
In his exposition of the hypothesis, Ball \citep{ball80-1} discusses whole spectrum of possible answers to the Fermi paradox, among them several variants of the Zoo hypothesis. The main variant, in which the Aliens (single species or some pan-species Galactic authority declared the Earth as strictly forbidden, may be weakened, so that there are a few ET scientists studying us  in some detail but inconspicuously; or allowing some occasional dabbling in our affairs; finally treating whole Earth ecosystem as an experiment in their laboratory, with multi-scale activities. What is common in these hypotheses is that the observation/activity is done by alien equivalents of our scientists, with the ultimate goal of knowledge and understanding.  Of course, the very name of the Zoo hypothesis suggests another type of Alien observers: public coming to view our planet and us as something amusing.

\section{Current hypothesis}

\subsection{Musings on a million year old civilization}
To present our hypothesis explaining the reported bizarre behavior of Aliens on our planet We shall start with some general considerations for the Alien civilization(s). In our discussion we shall concentrate on civilizations built by individualistic creatures. We argue that `single entity', `hive mind' and similar, deeply integrated societies, while possible, would play a less important role in deep space exploration. As our overall assumption is that the Universe is governed by physical laws, to which our current knowledge is a reasonably good approximation, we do not take into account phenomena such as faster-than-light space travel, instantaneous communication etc. In such a case, the physical separation of the interstellar scale would mean breaking contact (or seriously slowing it down) between physically separated members and groups of members of such hive-mind civilization. In our opinion this would pose a significant `psychological' barrier against separation required by space exploration.

Such barrier is absent in civilizations composed of beings with individual minds, accustomed to partial communication and independence of actions. But even here some general limitations would apply. 

Let us consider a million year old civilization. What can we deduct about its relevant characteristics? First: to survive for so long it must have achieved some sort of stability, against external factors (such as lack of resources, cosmic catastrophes, entropic factors etc.) as well as with respect to internal problems. Projecting humanity problems we would put internal conflict at the top of these internal issues, but this may be simply our local idiosyncrasy. What is more probable as the cause of the existential stress is boredom. First, to build a civilization, the intelligent beings must posses some amount of curiosity, to search for new solutions, to discover. And with curiosity fulfilled comes boredom. 
Such boredom would have tremendous effect on development of sciences and scientific interests. 
A few years ago, Horgan \citep{horgan96-1} has prophesied a quick end to sciences in our civilization, simply because we would run out of truly new things to discover. This was written only circa 3000 years after the first scientific efforts, and only 400 years after the start of the scientific revolution. While we may treat such statements as premature in our case, it is quite reasonable to expect that an advanced civilization may really discover all that is to be discovered `quite early' in their history, say in the first 100~000 years of their million-year lifetime. For the remaining period, the famous Rutherford's maxim would be viciously applicable: with no room for new physics, all science would be stamp collecting.

What then would be the subject that could keep the interest of such civilization alive? We think the answer is simple: the only structure rich enough and variable enough to sustain the curiosity is the civilization itself. Interactions between its members may grow rich beyond our speculative powers. 
We extrapolate here the explosive growth of dedicated entertainment and social networks seen in the past half century, from TV to multiple cable channels, to Internet and eventually to all the Web 2.0 ideas. Should our civilization survive the next few thousand of years, the flexibility of social communication  is truly hard to imagine. The interactions would grow much faster than the capabilities of individual humans, providing each with the environment fitting the individual needs. We think that this simple extrapolation is applicable not only to our society, but to most `interacting individuals' based extraterrestrial ones.  
And we should look at these interactions not through the scientific microscope of what we call psychology and sociology, which are but efforts to simplify. To keep the interest alive, the members of such advanced civilization would have to grok all the changing particulars, much more like literature than scientific point of view.
Even when we look at our own global civilization, with its $\sim10^{10}$ people we see that the majority of interest is in other human beings. Directly, via social structures, friends, lovers, enemies; or indirectly, with the proxies of mass media, literature, music, art, etc. Culture is  defined by people and their interactions rather than necessities of external, physical circumstances. Extrapolating this image to the imaginary extraterrestrials, we may assume only increased internalization of their cultural focus. 

Of course this does not mean that such civilization would lose all interest in `external' world. The security considerations alone would force it to develop and maintain technologies capable of monitoring cosmic environment and suitable reactions to dangers. Moreover, we can assume that during the million year development, the Aliens would master their environment sufficiently to allow enough resource reserves to fulfill the needs of its members.

\subsection{Trillion-channel TV}

Would such civilization actually explore the Universe, in a way that our more optimistic SF writers propose? Would they send organized exploration expeditions, or mercantile missions or even invasion fleets? We seriously doubt it. Resorting again to analogy with our society, we observe that with increased accessibility of trivial (but culture based!) content and communication the interest in exploring the outside world are not on the rise -- just the opposite. Popular media stars (starlets?) have much more impact on social mind than space exploration. In the unimaginably more complex world of internal relationships of the million year old civilization this ratio would be even smaller. 
Our guess is that they would keep what is a must to preserve their security, but, in general, no more than that.

We should now distinguish two cases. The first is that the increase of complexity of civilization would be achieved via increased capabilities of individual members with relatively small numbers (civilization of `a few sages', where `a few' might mean anything from a few to millions) and truly large society (extending well beyond our own numbers, i.e. more than $10^{10}$). We are especially interested in the latter case.
Billions of aliens, who, as we assumed are individuals, would differ in their characteristics: capabilities, histories, interests. Statistical laws would lead to conclusion that while the alien society as a whole might not be focused on exploring the Universe, some individuals would be interested in such activities. Assuming that their civilization is `rich enough', there might be also resources and technical solutions allowing such particular members interstellar travel. While we have excluded faster than light communication from our consideration, we assume that during long span of the civilization the control over life processes allowing to travel over the galaxy within single lifespan (hibernation, slowing down of life processes etc.) would be certain. It is our conclusion, therefore, that from the existence of a million year old civilization, even extremely introspective, we can expect some small fraction of `space travelers'. In such a case they would be rather from fringes of the specific social `norm', rather than the core of society. One of the reasons for this assumption is that galactic travelers would inevitably lose contact with their own civilization. This phenomenon has been present in numerous SF works: traveling to and from a location distant by mere few hundred light-years means that while the involved individual might age reasonably little (due to relativistic effects, hibernation or both), on his home solar system, hundreds of years would have passed. If we can judge by our own example, this would create unbridgeable  gap between the individual and the society. To decide on such `one way' trip, the space traveling Aliens might truly be judges asocial (within their standards\ldots).

Let us use an analogy from human society here. The development of cable and satellite TV has dramatically increased the amount of content broadcast to our societies. In some countries the number of available TV channels is now well over 200. For example, in the UK, Broadcasters' Audience Research Board (BARB, \url{http://www.barb.co.uk)} publishes audience data on over 270 TV channels. The results are very interesting: in addition to a few popular channels (with viewing share over 10\%) there are many channels for which BARB states that the viewer figures are small but not zero. 

Imagine now a trillion-channel `entertainment' environment for our advanced civilization (necessary to keep the curiosity alive). Most of the `channels' would be related to internal activities of the civilization. But a small fraction, say one millionth, might be `devoted' to the more interesting phenomena in the galaxy: spectacular cosmic events, such as supernova explosions, planets harboring life and possibly `low level' civilizations, such as ours. A few `viewers' would be sprinkled among these `channels'.

Continuing the analogy, the low popularity channels are typically special interest ones, catering to the needs of specific audiences. Let us, for a moment, consider what our planet has to offer to galactic viewers. Certain amount of violence, both on individual and massive scale might be interesting (compare this to us watching TV programs on wars between ants or feeding habits of lions). Intricacies of our mating habits might be interesting as well. We could treat these conclusions as extension of the Zoo hypothesis: we are interesting (for some Aliens) because of our, not too pleasant, characteristics as species (and possibly, whole ecosystem). Let us quote J. A. Ball \citep{ball80-1}: 
`\textit{As working hypotheses, I suggest, first,
that mankind is neither alone nor number one. Advanced civilizations exist
and exert some degree of control over the galaxy. Second, they're aware
of us at least at some level. But are they concerned with us? We may be
only an obscure entry in their tabulation of inhabited regions of the galaxy.}' 
But where we differ from Ball is in the vision of who comes to observe us (and maybe meddle a little in our affairs).

We come then to a natural solution of the question why the Aliens, reported in popular accounts engage in behaviors such as abductions, often performed with the aim of sexual experimentation, senseless aggression (such as cattle mutilations) or even helping local cultures by advanced technologies which make wars more intense and bloody (and additionally give the Alien(s) status of deities). Thus what Eberlein \citep{eberlein01-1} has called the `\textit{naive-realistic}' hypothesis, is perhaps not so naive after all. Maybe, on the contrary, it is naive to expect, idealistically, that the visitors coming to our little blue planet would be the `cream of the cream' of the Alien civilization: scientists, military leaders etc. 
If the visitors are scientists -- then it would be a very special kind of scientist, interested in such low complexity subject as ourselves. And as for interstellar diplomatic relationships: would we consider sending diplomats to a dung beetle?
We think that at least some of the multitude of bizarre reports are true: this is the kind of Aliens that are attracted to Earth. Singly, or in small groups, they have been present among us, with our planet being a decidedly minor channel in the galactic  interactive entertainment network. Like channel, like viewers: for what interesting aspects the Earth can offer, we get as visitors murdering predators, sexual deviants, mad scientists or at best, petty dictators. 

Should we keep the metaphor proposed by the Zoo hypothesis, perhaps our part of the Universe is not a well maintained and constantly watched and managed zoo but rather a wilderness preservation zone, off-limits to law-abiding Galaxy citizens. And, as with our Earthly counterparts, one might find a few trespassers (corresponding to illegal hunters, convicts on the run, extreme environmentalists or survival fans).

We postulate that the aliens visiting Earth would, most likely, represent the far tail of the Gaussian distribution of their species, too stupid to comprehend/appreciate the intricacies of their own culture, but bright enough to feel satisfaction from meddling in ours.
Consider, for example, the accounts of aliens living among us. Just what kind of a representative of advanced extraterrestrial civilization would like to change their biology (or to invade the brain) to successfully hide its identity and  live as a member of our society?
The solution to the Fermi paradox is trivial: there is Alien life in the Galaxy -- and it has been observed on Earth. At least some of the reports of the close encounters, strange as they may seem, may be true. The problem is that 
the scientists do not believe them. This requires change of attitude. 

There is a perfect counterexample that provides some insight as to the way we should approach the situation. This is the early research on AIDS, or `gay cancer' as it was called in 1981. Despite rarity of the cases and social stigmatization, the Communicative Diseases Center has quickly formed a specialized task force to coordinate the research. This was done despite the fact that some of the key CDC people did not believe that there is a great danger, clearly stating in public that there is no risk outside the male homosexual community \cite{altman81-1}. 
Yet the research continued, in opposition to social branding of the disease, to discover the full scope of the risk and danger to groups thought previously to be totally safe. We would like to think that everything was done as it should. Still, in his review of social and professional approach to AIDS appearance, Eisberg \cite{eisenberg86-1} has written `\textit{One might be tempted to conclude that questions of public health policy are best dealt with by expert judgment unsullied by lay opinion. Yet such an attitude supposes scientists to be governed by pure reason and to be beyond influence by narrow self-interest or by political and moralistic considerations, a self-serving assumption belied by an examination of the record.}'

The UFO case is clearly an opposite: while significant part of general public believes in the reality and importance of the Alien encounters, the scientists, by and large, refuse to take action other than negatively branding the people who report such events. Which interests are served here? Especially in terms of risk management?
Instead of treating people who report meeting Aliens as mental cases, we (the scientists) should work on a much more disquieting hypothesis, that the visitors from outer space, are more likely to be abnormal minorities (by the standards of their own civilizations) than scientists, diplomats or even military commanders. Maybe, even if we do not discover the Aliens themselves,  we would find some statistical clues to their numbers and intentions. Serious approach to close encounter accounts is reasonable strategy, even when we think that the probability is very low, the possible damage is huge, so the risks may be high \cite{kaplan81-1}. Because we do not know the capacities for destructive power in the hands (tentacles?) of these mad visitors, we should be afraid, be very afraid.

\section*{Acknowledgments}
I would like to thank Mr Ksawery Stojda for fruitful discussions and all the unknown monsters from outer space for not kidnapping nor mutilating me (so far).
This work has not been supported by any grant whatsoever.

\section*{Post publication note, July 2012}
The train of thought explored in this work is hardly original, although probably not obviously so. Similar notions were considered before, in much serious contexts. For example, on July 12, 2012, Britain's  Ministry of Defence's (MoD) unclassified and published document archives which revealed that the MoD staff believed that aliens could visit the Earth for ``military reconnaissance", ``scientific" research or ``tourism". Specifically, 
in a 1995 briefing by a UFO desk officer at DI55 has drawn the attention to the need for a full study of UFO data as national security, as the implications have never been properly assessed.  He stated ``\textit{there is no hard evidence for alien craft}" but  ``\textit{if the sightings are not of this earth then their purpose needs to be established as a matter of priority. There has been no apparently hostile intent and other possibilities are: a) military reconnaissance b) scientific c) tourism}''. 

This notion itself is also not original, as it happened fifteen years after Douglas Adams' Hitch Hikers Guide to the Galaxy was broadcast and published\ldots

\end{document}